\documentclass[10pt,twocolumn,twoside]{IEEEtran}

\usepackage{setspace} 
\usepackage{graphicx}        
\usepackage{amssymb}
\usepackage{graphicx}
\usepackage{algorithm}
\usepackage{algpseudocode}
\usepackage{amsmath,todonotes}
\usepackage{mathtools}
\usepackage{comment}
\usepackage{multirow}
\usepackage{booktabs}
\usepackage[noadjust]{cite}
\usepackage{balance}
\newcommand\x{\times}

\newtheorem{problem}{Problem}
\newtheorem{proposition}{Proposition}
\newtheorem{lemma}{Lemma}
\newtheorem{remark}{Remark}

\newtheorem{assumption}{Assumption}
\newtheorem{theorem}{Theorem}

\newtheorem{example}{Example}

\begin{document}
%

\title{\LARGE \bf Estimating the Impact of Cyber-Attack Strategies for \\ Stochastic Control Systems}

\author{Jezdimir Milo\v{s}evi\'{c}, Henrik Sandberg, and Karl Henrik Johansson$^{1}$
\thanks{*This work was supported by the Swedish Civil Contingencies Agency through the CERCES project, the Swedish Research Council, Knut and Alice Wallenberg Foundation, and the Swedish Foundation for Strategic Research.}
\thanks{$^{1}$The authors are with the Department of Automatic Control, School of Electrical Engineering and Computer Science, KTH Royal Institute of Technology, Stockholm, Sweden. Emails:\{jezdimir, hsan, kallej\}@kth.se.
}
}

\markboth{}
{Shell \MakeLowercase{\textit{et al.}}: Bare Demo of IEEEtran.cls for IEEE Journals}

\maketitle

\begin{abstract} 
Risk assessment is an inevitable step in implementation of a cyber-defense strategy. 
An important part of this assessment is to reason about the impact of possible attacks.
In this paper, we propose a framework for estimating the impact of cyber-attacks in stochastic linear control systems. 
The framework can be used to estimate the impact of denial of service, rerouting, sign alternation, replay, false data injection, and bias injection attacks. 
For the stealthiness constraint, we adopt the Kullback-Leibler divergence between residual sequences during the attack. 
Two impact metrics are considered: 
(1) The probability that some of the critical states leave a safety region; and
(2) The expected value of the infinity norm of the critical states. 
For the first metric, we prove that the impact estimation problem can be reduced to a set of convex optimization problems. 
Thus, the exact solution can be found efficiently. 
For the second metric, we derive an efficient to calculate lower bound.  
Finally, we demonstrate how the framework can be used for risk assessment on an example.
\end{abstract}

\section{Introduction}
Industrial control systems (ICS) operate physical processes of  great societal significance.
Examples include electricity production, water purification, transportation, and oil and gas distribution. 
For this reason, ensuring security of ICS is very important.
Unfortunately, it is known that many security vulnerabilities can be found within these systems~\cite{nist,msb}.  
These vulnerabilities can be exploited by malicious attackers to harm ICS, as demonstrated through several incidents that have happened so far~\cite{Slay2008,stuxnet,case2016analysis}. 
Therefore, effective cyber-defense strategies have to be developed. 

However, protecting ICS is more complicated and costly than protecting regular IT systems~\cite{cardenas2008research}. 
Thus, the recommendation is to conduct a risk assessment prior to implementing a defense strategy~\cite{nist,msb}. 
The goal of this assessment is to prioritize among vulnerabilities, which is done based on the likelihood that vulnerabilities are exploited, and the impact that can happen if exploitation occurs~\cite{stoneburner2002sp}. 
The resources can then be focused in preventing the most critical vulnerabilities.

Motivated by the risk assessment application, we study the impact estimation problem.
This problem reduces to a constrained maximization problem, where the objective function is an impact metric. 
The constraints of the problem vary, but in most of the cases they include stealthiness constraint. 
In other words, we want to check if the attacker can conduct large damage and remain stealthy at the same time.


So far, significant effort has been dedicated towards estimating impact of attacks that remain undetected by the \textit{chi-square} anomaly detector~\cite{5718158,carlos_IFAC,jovanov_2017,moura_2017,7470566,jezdimir_bias}.
In these studies, the reachable sets were predominantly used to characterize the impact, and algorithms for calculating upper and lower bounds of these sets were proposed in several works~\cite{5718158,carlos_IFAC,jovanov_2017}.  
Other types of detectors have also been considered~\cite{cardenas,8431073,Urbina2016,ahmed2017model,murguia2018security,david}.
For example, C{\'a}rdenas \textit{et al.} considered a cumulative sum (CUSUM) detector, and investigated the impact of several attack strategies that remain undetected by this detector~\cite{cardenas}. 
In~\cite{8431073}, the impact of zero alarm attacks was used to compare chi-square, windowed chi-square, and CUSUM detectors. 

Two directions in which the aforementioned studies can be extended are outlined next. 
Firstly, these works usually assume resourceful attackers, with the full model knowledge and the ability to inject arbitrary signals to the sensors and actuators it controls.
However, less complex attacks, which are also more likely to happen, are  important to consider for risk assessment. 
For instance, we may encounter the case where a certain vulnerability allows the attacker to deny the operator access to a signal, but not to set the signal to an arbitrary value. 
Moreover, comparing vulnerabilities based on a single attack model may lead to a biased judgment. 
Secondly, these works focus their analysis on particular types of anomaly detectors.
Thus, the impact analysis is carried out for every detector separately. 
In summary, a more general framework for risk assessment is desirable.

The main obstacle towards such a framework is the nature of the worst-case impact estimation problem, which generally is not possible to formulate as a convex optimization problem.
Exceptions do exist, however. 
For example, in~\cite{david}, the authors introduced the infinity norm of critical states to quantify the impact under the CUSUM detector.
They showed that the exact value of impact can be obtained by solving a set of convex problems. 
This useful property of the infinity norm was also recognized in~\cite{hirzallah2018computation,Milosevic_workshop}, where the impact was obtained by solving a set of linear programs. 
However, these works neglect the influence of noise. 
Thus, to extend the analysis to more general stochastic systems, a suitable replacement for the infinity norm with similar convenient properties is needed.

The contributions of our work are as follows. 
Firstly, we propose a unified framework for quantifying the impact of cyber-attack strategies for stochastic control systems. 
The framework covers both advanced injection strategies such as false data injection (FDI)~\cite{5718158,david} and bias injection~\cite{ahmed2017model,bias}, but also simpler strategies such as denial of service (DoS)~\cite{cardenas2008research,amin2009safe}, replay~\cite{6612700}, rerouting~\cite{teixeira2015voltage,ferrari2017detection},
and sign alternation attacks~\cite{7470566}.
Additionally, Kullback-Leibler (KL) divergence~\cite{cover2012elements} is used to define a stealthiness constraint, which makes the framework independent of the choice of anomaly detector. 
Stealthiness constraints based on KL--divergence have been used in several works so far~\cite{GUO2018117,7947172,7799044,7466835}. 

The second and third contributions concern solving the impact estimation problem. 
Particularly,  we propose two metrics as a substitute for the infinity norm in stochastic systems:
The probability that some of the critical states leave a safety region ($I_1$) and the expected value of the infinity norm of the critical states ($I_2$).  
As the second contribution, we show that $I_1$ has the same desirable properties as infinity norm.
That is, we prove that the exact value of the impact for this metric can be obtained by solving a set of convex problems (Theorem~1). 
Unfortunately, $I_2$ does not have the closed form expression in general, and is not trivial to evaluate. 
However, we show how to obtain an efficient to calculate lower bound for $I_2$ (Theorem~2), which is our third contribution.    
Finally, the applicability of our framework for risk assessment is demonstrated on a numerical example. 

We remark that the preliminary version of this paper appeared in~\cite{jezdimir_ECC}. This work differs from~\cite{jezdimir_ECC} in the following aspects: (1) A stochastic model is considered; (2) This work is not constrained to certain classes of anomaly detectors; (3) Different impact metrics are used; (4) Due to the differences in {(1)--(3)}, a different theoretical analysis is required to derive the main results; (5) The numerical example is extended.

The paper is organized as follows. 
In Section~\ref{section:system_description}, we introduce the model setup. 
In Section~\ref{section:impact_estimation}, we formulate the impact estimation problem. 
In Section~\ref{section:main_results}, we present the main technical results of the paper. 
In Section~\ref{section:attack_strategies}, we introduce attack strategies compatible with our framework.
In Section~\ref{section:simulations}, we illustrate the applicability of our framework on a numerical example.  
In Section~\ref{section:conclusion}, we conclude. 

\textit{Notation. }We denote by: $\textbf{0}_{m\times n}$ the zero-matrix with $m$ rows and $n$ columns; $\textbf{I}_n$ the identity matrix with $n$ rows and columns; $T(i,:)$ the $i$-th row of matrix $T$; $T(i,j)$ the element of matrix $T$ positioned in the $i$-th row and $j$-th column; $f^{(i)}$ the $i$-th element of vector $f$; $\mathcal{N}(\mu,\Sigma)$ the Gaussian distribution with the mean $\mu$ and the covariance $\Sigma$. For a signal $x: \mathbb{Z}_{\geq 0} \rightarrow \mathbb{R}^n$, $x_{0:N}$ is given by $x_{0:N}=[x(0)^T \ldots x(N)^T]^T$.


\section{Model Setup\label{section:system_description}} 


%



The system consists of a physical plant, estimator, controller, and anomaly detector. 
The plant is modeled as
\begin{equation}
\begin{aligned}
x(k+1)&=Ax(k)+B\tilde{u}(k) + v(k) \\
y(k)&=Cx(k) + w(k)
\label{eqn:systemg}
\end{aligned}
\end{equation}
where $x(k) \in \mathbb{R}^{n_x}$ is the plant state, $y(k) \in \mathbb{R}^{n_y}$ is the vector of measurements, $\tilde{u}(k)\in \mathbb{R}^{n_u}$ is the control signal applied to the plant, and  $v(k) \in \mathbb{R}^{n_x}$ and $w(k) \in \mathbb{R}^{n_y}$ are process and measurement noises, respectively. 
These processes are independent Gaussian white processes with zero mean and covariance matrices $\Sigma_v$ and   $\Sigma_w$. 
The initial state $x(-\infty)$ is zero mean Gaussian with covariance $\Sigma_{-\infty}$ and independent of $v$ and $w$.
We assume $\Sigma_v$, $\Sigma_w$, $\Sigma_{-\infty}$ are positive definite, the pair $(C,A)$ is observable, and the pair $(B,A)$ is controllable.

%




To estimate the state $x$, the Kalman filter is used. We assume that the filter has converged to the steady state regime before the attack starts, so the state estimate is given by
\begin{equation}\label{eqn:estimate}
\begin{aligned}
\hat{x}(k+1)=(A-KC)\hat{x}(k)+Bu(k)+K\tilde{y}(k)
\end{aligned}
\end{equation}
where $\hat{x}(k) \in \mathbb{R}^{n_x}$ represents the one step ahead prediction of $x(k)$, $u(k) \in \mathbb{R}^{n_u}$ is the control signal calculated by the controller, and $\tilde{y}(k)\in \mathbb{R}^{n_y}$ is the measurement signal received by the estimator. The steady state Kalman gain is given by $K=A\Sigma_e C^T(C\Sigma_eC^T+\Sigma_w)^{-1}$
where $\Sigma_e$ is the error covariance matrix obtained by solving the Riccati equation $\Sigma_e=A\Sigma_eA^T+\Sigma_v-A\Sigma_eC^T(C\Sigma_eC^T+\Sigma_w)^{-1}C\Sigma_eA^T. $
This gain exists under the assumptions on the noise processes, and it is also known that $A-KC$ is asymptotically stable~\cite{optimalfiltering}.


The controller is of the form
\begin{equation} \label{eqn:controller}
u(k)=-L_{\hat{x}} \hat{x}(k)+L_{y_r}y_r(k)
\end{equation}
where $y_r(k) \in \mathbb{R}^{n_{y_r}}$ is a bounded reference signal that satisfies
\begin{equation} \label{equation:referenc_constraitns}
||Q_{y_r} y_r(k)||_{\infty} \leq 1
\end{equation} 
and $Q_{y_r} \in \mathbb{R}^{n_{y_r} \times n_{y_r}}$ is a full rank scaling matrix. The direct term $L_{\tilde{y}}\tilde{y}(k)$ can also be added to $u(k)$, but we omit it for the sake of brevity.  
We assume that stability and satisfactory performances are obtained in absence of attacks. 
Additionally, given that the reference is usually a constant signal that is not updated often, we adopt the following standing assumption.
\begin{assumption}\label{assumption:constant_reference}
The reference signal is a constant signal $y_r(k)=y_r$ that does not change during the attack. The system has reached stationary regime before the attack starts.  \hfill $\square$
\end{assumption}


In absence of attacks, $\tilde{y}(k)=y(k)$ and $\tilde{u}(k)=u(k)$. 
However, because of an attack or a fault in the system, these values may differ. 
To detect these anomalies, an anomaly detector is used. 
The first step of detection procedure is to calculate the residual signal
\begin{equation}\label{eqn:residual}
\tilde{r}(k)=\Sigma_r^{-1/2}\big(\tilde{y}(k) - C\hat{x}(k) \big)
\end{equation}
where $\Sigma_r= C\Sigma_eC^T+\Sigma_w.$ 
Note that $C\hat{x}(k)$ is the estimate of the measurement vector based on the previous measurements and the system model.
Thus $\tilde{r}(k)$ represents the scaled difference between $\tilde{y}(k)$ and its modeled (expected) behavior. 
In the absence of anomalies, the residual sequence is a \textit{white}  Gaussian process with \textit{zero} mean value and \textit{identity} covariance matrix. 
Throughout the paper, we denote by $r$ the non-attacked residual signal to distinguish it from the attacked one. 

The second step of the detection procedure is to process the residual signal $\tilde{r}$ to obtain a security metric $S$. When this metric exceeds a certain predefined threshold, an alarm is raised. 
How $S$ is determined depends on the detector. 
Some examples of the detectors include chi-square~\cite{1104658}, CUSUM~\cite{basseville1993detection}, and multivariate exponential weighted moving average~\cite{1023071269551} detectors.
However, in this work, we are not focused on any particular anomaly detector, so we do not need to define $S$.

We now introduce the model of attacker. 
For simplicity, we assume that the attack starts at $k=0$. 
The measurements $\tilde{y}(k)$ and the control signals $\tilde{u}(k)$ are modeled by
\begin{equation}\label{eqn:attack} 
\begin{aligned}
\tilde{y}(k)&= \Lambda_{y}  y(k)+\Gamma_{y} a_y(k)+\Gamma_{y}a_{s}(k) \\
\tilde{u}(k)&=\Lambda_{u} u(k)+\Gamma_{u} a_u(k) 
\end{aligned}
\end{equation}
where $a_u(k) \in \mathbb{R}^{n_{a_u}}$ is the deterministic part of the attack against the actuators, $a_y(k) \in \mathbb{R}^{n_{a_y}}$  is the deterministic part of the attack against the sensors, and $a_s(k)\in \mathbb{R}^{n_{a_y}}$ is a stochastic signal the attacker injects.
The matrices $\Gamma_{y}$, $\Gamma_{u}$, $\Lambda_{y} $, and $\Lambda_{u}$ depend on the attack strategy and the attacker's resources. 
We explain how these matrices are formed in Section~\ref{section:attack_strategies}.
%


By combining equations \eqref{eqn:systemg}--\eqref{eqn:controller}, \eqref{eqn:residual}, \eqref{eqn:attack}, the dynamics of the system under attack can be expressed as
\begin{equation}\label{eqn:system_extended_eq_at}
\small
\begin{aligned}
x_e(k+1)&=\tilde{A}x_e(k)+\tilde{B} f(k)+\tilde{E} y_r+\tilde{G} a(k)+\tilde{J} a_s(k)\\
\tilde{r}(k)&=\tilde{C} x_e(k)+\tilde{D} f(k) +\tilde{F} y_r+\tilde{H} a(k)+\tilde{K} a_s(k)\\
\end{aligned}
\end{equation}
where $x_e(k)=[x(k)^T \hspace{1mm} \hat{x}(k)^T]^T$, $f(k)=[v(k)^T \hspace{1mm} w(k)^T]^T$, and $a(k)=[a_u(k)^T \hspace{1mm} a_y(k)^T]^T$ are the extended state, noise, and attack vectors, respectively. We denote the dimension of $a(k)$ by $n_a$, and of $f(k)$ by $n_f$. 
 The extended system matrices are
\begin{align*}
\tilde{A} &= 
\begin{bmatrix}
A &  -B\Lambda_u L_{\hat{x}}               \\
K\Lambda_yC  & A-KC-BL_{\hat{x}}       
\end{bmatrix}
\hspace{-2mm}&&\tilde{B} =
 \begin{bmatrix}
\textbf{I}_{n_x}  & \textbf{0}_{n_x \times n_{y}} \\
\textbf{0}_{n_x \times n_{x}}   &      K \Lambda_y
\end{bmatrix}
\\
  \tilde{C} &=\Sigma_r^{-1/2}\begin{bmatrix}
\Lambda_yC  & -C
\end{bmatrix} 
&&\tilde{D}=
[\textbf{0}_{n_{y} \times n_x}    \hspace{1mm}    \Sigma_r^{-1/2}\Lambda_y]
\\
\tilde{E}&= 
\begin{bmatrix}
B \Lambda_u L_{y_r}        \\
BL_{y_r}    
\end{bmatrix} 
&&\tilde{F} =
\textbf{0}_{n_{y} \times n_{y_r}}
\\
\tilde{G}&= 
\begin{bmatrix}
B\Gamma_{u}  &  \textbf{0}_{n_x \times n_{a_y}}  \\
\textbf{0}_{n_x \times n_{a_u}}    &      K\Gamma_y
\end{bmatrix}
 &&\tilde{H} = 
[\textbf{0}_{n_y \times n_{a_u}}  \Sigma_r^{-1/2} \Gamma_{y}]
\\
\tilde{J}&= 
\begin{bmatrix}
\textbf{0}_{n_x \times n_{a_y}} \\
K\Gamma_y
\end{bmatrix} 
 &&\tilde{K}=   
 \Sigma_r^{-1/2} \Gamma_{y}.
\end{align*}

\section{Problem Formulation} \label{section:impact_estimation}

In this section, we formulate the problem of estimating the attack impact. 
Prior to that, we introduce the decision variables, the impact metrics, and some of the constraints.

\subsection{Decision Variables}

We are interested in estimating the impact of an attack on the subset of states $x$ during $N$ time steps.
Additionally, we want to check if the attack remains stealthy during this time interval.
From~\eqref{eqn:system_extended_eq_at}, $x$ is influenced by noise $f$, reference $y_r$, and attack signals $a$ and $a_s$. 
As shown later, the impact metric and the constraints are not dependent on a particular realization of sequences  $f$ and $a_s$.
However, they do depend on $y_r$ and $a_{0:N}$.
Since we perform off-line analysis of the attack impact, we do not know the exact value of the reference at the beginning of the attack. 
Thus, we identify the worst possible value of $y_r$ when estimating the impact. 
We also assume that the attacker selects the worst possible sequence $a_{0:N}$ to maximize the damage. 
Thus, we define the decision variables of the problem to be $y_r$ and $a_{0:N}$, and denote them by
$d=\begin{bmatrix} a_{0:N} \\ y_r \end{bmatrix}.$


\subsection{Impact Metrics}
The impact metrics we define are based on the concept of critical states. These states are defined as $z(k)=Q_z x(k)$,  
where $Q_z  \in \mathbb{R}^{n_z \times n_x}$ is a full row rank scaling matrix, and  $n_z\leq n_x$ is the number of the critical states. 
The critical states may model the flow of energy through the power line that should be maintained within predefined bounds, or a temperature that should not exceed some safety limit. 
If any of the critical states is larger than $1$, that indicates a dangerous state of the system.

\begin{example}
Let $x=[x^{(1)} \hspace{1mm}x^{(2)}]^T$. 
Let $x^{(2)}$ be the critical state that should be kept within the interval $[-\bar{x}^{(2)},\bar{x}^{(2)}]$, $\bar{x}^{(2)} \geq 0$. The matrix $Q_z$ is then defined by $Q_z=[0 \hspace{2mm} 1/\bar{x}^{(2)}]$. 
Therefore, if $|x^{(2)}(k)| \geq \bar{x}^{(2)}$, then $|z(k)| \geq 1$. 
 \hfill $\square$
\end{example}


In the previous work on deterministic systems~\cite{david}, the impact metric was defined as 
\begin{equation} \label{eqn:deterministic_metric}
I(d)=||z_{1:N}||_{\infty}.
\end{equation}
Values of $I$ larger than 1 imply that the attacker can drive some of the critical states outside the safety region. 
However, in this paper, the state is influenced by the noise in addition to attacks. 
Therefore, even in the absence of attacks, some of the critical states can become larger than 1 with non-zero probability.
To make the impact metric more suitable for stochastic systems, we define a new metric 
\begin{equation}\label{equation:impact_stochastic}
I_1(d)= \text{max}_{i \in \{1,\ldots,n_z N\}} \hspace{2mm} \mathbb{P}(|z^{(i)}_{1:N}| \geq 1). 
\end{equation}
If $I_1$ is close to 1, then we know there is a high probability that some of the critical states leave the safety region.
If $I_1$ is close to 0, the critical states stay within the safety region with high probability, and the attack is harmless.
Another possible impact metric is
 \begin{equation}\label{eqn:impact_stochastic_alternative}
I_2(d)=\mathbb{E}\{||z_{1:N}||_\infty\} .
 \end{equation}
Unfortunately, the expected value of the $\infty$-norm does not have a closed form expression, and is hard to evaluate in general. 
Thus, we are primarily focused on the metric $I_1$.

\subsection{Stealthiness Constraint and Constraints on Attack Signals}

We first discuss the stealthiness constraint. 
 To define this constraint, we use KL--divergence. 
The KL--divergence gives the distance between probability density functions $p$ and $q$ over a sample space $X$, and is defined by
$$\mathcal{D}(p||q) = \int_{X} \log \bigg(\frac{p(x)}{q(x)}\bigg) p(x) dx.$$ 
It is known that $\mathcal{D}(p||q)\geq 0$ with equality if and only if $p$ equals $q$ almost everywhere. 
The KL--divergence stealthiness constraint can take different forms~\cite{GUO2018117,7947172,7799044,7466835}. 
Since we observe the attack over the time interval $[0,N]$, we adopt the constraint
\begin{align} \label{eqn:stealthiness}
\frac{1}{N+1}\mathcal{D}(\tilde{r}_{0:N}||r_{0:N}) \leq \epsilon.   
\end{align}
The idea is that if the KL--divergence between the probability density functions of attacked $\tilde{r}_{0:N}$ and non-attacked $r_{0:N}$ residual sequences remains small, then these density functions are similar, and the attack stays stealthy. 

The advantages of using KL--divergence as a stealthiness constraint include the following. 
Firstly, as shown later, this constraint is a convex symmetric constraint for the class of attacks that we observe. 
Secondly, the impact analysis is made independent of the choice of anomaly detector. 
Thirdly, generating attacks that satisfy~\eqref{eqn:stealthiness} can be a reasonable choice by the attacker that does not know which anomaly detector is deployed. 
Finally, it was shown that for certain types of attacks and anomaly detectors, other types of stealthiness constraints can be replaced by a KL--divergence based constraint~\cite{5718158}.

The second set of the constraints we discuss come from a particular attack strategy. 
In general, we do not know which attack strategy the attacker will use.  
Thus, to reason about the attack and estimate its impact, a number of strategies has been modeled throughout the literature. 
In this paper, we consider DoS, rerouting, sign alternation, replay, FDI, and bias injection attacks.
Each of these strategies introduces additional constraints on $a$ and $a_s$. 
However, we show in Section~\ref{section:attack_strategies} that all of the strategies can be captured by~\eqref{eqn:attack},  and the constraints
\begin{align}
F_a a_{0:N}&=0  \label{eqn:constraint_a} \\
a_{s0:N}&=T_{sx}x_e(N_s)+T_{sr} y_r+T_{sf} f_{N_s:-1} \label{eqn:as} 
\end{align}
where $N_s<0$, $F_a \in \mathbb{R}^{n_{F_a} \times (N+1) n_a}$, and $T_{sx}$, $T_{sr}$, and $T_{sf}$ are the matrices of appropriate dimensions. 




\subsection{Problem Formulation}
Based on the previous discussion, we introduce the problem of estimating attack impact. We first consider the metric $I_1$. 
\begin{problem}\label{problem:impact_estimation} Estimating impact using $I_1$ 
\begin{align*}
\underset{d}{\text{maximize } }\hspace{4mm} & I_1(d)  \\
\text{subject to } \hspace{4mm} 
&x_e(k) \text{ and } \tilde{r}(k) \text{ satisfy~\eqref{eqn:system_extended_eq_at}} &&\text{(C1)}\\
& a_{s0:N}\text{ satisfies~\eqref{eqn:as} } &&\text{(C2)}\\
  &||Q d||_{\infty} \leq 1 &&\text{(C3)}\\
&\frac{1}{N+1}\mathcal{D}(\tilde{r}_{0:N}||r_{0:N}) \leq \epsilon  &&\text{(C4)}\\
 &F d =0 &&\text{(C5)}
     \end{align*} 
     \end{problem}
In this problem, we check if the attack can have large impact on the system by maximizing $I_1$.
The constraint: (C1) ensures that physical equations of the system are satisfied; (C2) imposes the assumption on the form of $a_{s0:N}$; (C3) comes from the reference constraint~\eqref{equation:referenc_constraitns}, where $Q=[\textbf{0}_{n_{y_r}\times(N+1)n_a} \hspace{1mm} Q_{y_r}]$; (C4) is the stealthiness constraint; (C5) comes from~\eqref{eqn:constraint_a}, where $F=[F_a\hspace{1.5mm}\textbf{0}_{n_{F_a}\times n_{y_r}}]$. 
For brevity of notation, we denote the constraints (C1)--(C5) with $\mathcal{C}$.   
We also introduce the impact estimation problem based on $I_2$. 

\begin{problem}\label{problem:impact_estimation_2} Estimating impact using $I_2$ 
\begin{align*}
  \text{maximize}_{d \in \mathcal{C}} \hspace{2mm} I_2(d)
       \end{align*} 
     \end{problem}
The following remarks are in place. 
\begin{remark}
$N$ and $\epsilon$ should be seen as tuning parameters. 
Naturally, we first want to discover attacks that result in high impact in short amount of time, without changing the residual distribution considerably. 
Thus, choosing small values of $N$ and $\epsilon$ is a good starting point for the analysis.
If the budget is large enough to prevent these attacks, we can start increasing the parameters to discover less dangerous attacks.  
\hfill $\square$
\end{remark}

\begin{remark}
One can also consider maximizing impact in $N_z$ steps and imposing stealthiness in $N_r\neq N_z$ steps. 
The case $N_z<N_r$ captures attacks that maximize damage in $N_z$ steps, and prevent the defender noticing this in additional $N_r-N_z$ steps. 
The case $N_z>N_r$ models ambush attacks~\cite{LIPP201644}, where the attacker stealthily prepares $N_r$ steps, and then launches a not necessarily stealthy attack in the remaining time. 
Although we focus on the case $N_r=N_z=N$, the analysis that follows can be extended to cover the aforementioned cases as well. \hfill $\square$ 
\end{remark}
\begin{remark}Problems~\ref{problem:impact_estimation} and~\ref{problem:impact_estimation_2} can be infeasible due to (C4). If that is the case, we define the impact to be $0$. \hfill $\square$
\end{remark}

Problems~\ref{problem:impact_estimation} and~\ref{problem:impact_estimation_2} are constrained maximization problems. Unfortunately, efficient algorithms for solving these type of problems are unknown in general. Nevertheless, in the next section, we propose an efficient way to obtain the optimal value of  Problem~\ref{problem:impact_estimation}, and a lower bound for Problem~\ref{problem:impact_estimation_2}.

\section{Main Results} \label{section:main_results}

In this section, we prove that the optimal value of Problem~\ref{problem:impact_estimation} can be obtained by solving a set of convex optimization problems (Theorem~\ref{theorem:solving_problem_1}). 
Furthermore, we show that in the process of solving Problem~\ref{problem:impact_estimation}, we also obtain a lower bound of Problem~\ref{problem:impact_estimation_2} (Theorem~\ref{prop:alternative_impact_metric}). 
Prior to proving Theorems~\ref{theorem:solving_problem_1} and~\ref{prop:alternative_impact_metric}, we introduce some technical results.

\subsection{Preliminaries}
We begin by introducing the two lemmas. 
 The first lemma determines distribution of the state $x_e$ prior to attacks.
  The second one states that for the class of the attacks we observe, the mean values of $z_{1:N}$  and $\tilde{r}_{0:N}$ are linear in $d$ and have fixed covariance matrices.
  The proofs of all the lemmas from this section are presented in Appendix.
\begin{lemma}
\label{lem:lemma_initial} Assume that attacks are not present and that the system has reached the stationary regime. Then $x_e$ is distributed according to $\mathcal{N}(T_{0} y_r, \Sigma_{0})$.
  \end{lemma}
  
  \begin{lemma}
\label{lem:critical_states_properties} 
In the presence of attacks, $z_{1:N} \sim \mathcal{N}(T_Z d,\Sigma_{Z})$ and $\tilde{r}_{0:N}\sim \mathcal{N}(T_Rd,\Sigma_{R})$, where $T_Z$, $T_R$, $\Sigma_{Z}$, $\Sigma_{R}$ are not dependent on $d$, and $\Sigma_{Z}$ is positive definite. 
  \end{lemma}
  
While it can be proven that $\Sigma_{Z}$ is positive definite, the same claim does not hold for $\Sigma_{R}$. 
In what follows, we assume that $\Sigma_{R}$ is positive definite. 
We later argue that the cases where $\Sigma_{R}$ is positive semidefinite are not of practical significance. 
 \begin{assumption} \label{assumption:positive_definite}
$\Sigma_{R}$ is a positive definite matrix.
\end{assumption}
Once Assumption~\ref{assumption:positive_definite} holds, the constraint~\eqref{eqn:stealthiness} is well defined and represents a convex and symmetric constraint in $d$.
   \begin{lemma}\label{lem:stealthiness_constraint_properties}
The constraint~\eqref{eqn:stealthiness} reduces to the constraint $$d^T T_R^T T_R d \leq \epsilon'$$ where $\epsilon'=(N+1)(2\epsilon+n_y)- \text{tr}(\Sigma_R) +\text{ln }\text{det}(\Sigma_R)$.
\end{lemma}

Note that if some of the eigenvalues of $\Sigma_R$ approach 0, then $\epsilon'$ approaches $-\infty$. 
The constraint~\eqref{eqn:stealthiness} then becomes impossible to satisfy, which justifies excluding the cases where $\Sigma_{R}$ is semidefinite from the analysis.    

We now introduce the problem that turns out to be crucial for solving Problem~\ref{problem:impact_estimation} and lower bounding Problem~2:
\begin{align} \label{eqn:alternative_problem}
\text{maximize}_{d \in \mathcal{C}}\hspace{2mm} &\mu_i(d)=\mathbb{E}\{ z^{(i)}_{1:N} \} 
     \end{align} 
where $i \in \{1,\ldots,n_zN\}$. 
Note that except for the objective that is different, the constraints and the decision variables are the same as those for Problems~\ref{problem:impact_estimation} and~\ref{problem:impact_estimation_2}.
This problem is reducible to a convex optimization problem, which implies that it can be solved efficiently, using well established algorithms.

\begin{proposition} \label{theorem:convexity_problem_2}
The problem given by~\eqref{eqn:alternative_problem}  is reducible to a convex optimization problem with symmetric constraints. 
\end{proposition}
\textit{Proof.} From Lemma~\ref{lem:critical_states_properties}, $\mathbb{E}\{ z_{1:N}^{(i)} \}=T_Z(i,:) d$. Thus, the objective function is linear in $d$, so it is simultaneously concave and convex. 
The constraints (C1) and (C2) enforce distributions of $z_{1:N}$  and $\tilde{r}_{0:N}$ given in Lemma~\ref{lem:critical_states_properties}.
The constraint (C3) is symmetric and convex in $d$ due to the infinity norm.  
From Lemma~\ref{lem:stealthiness_constraint_properties}, the constraint (C4) can be exchanged with $d^T T_R^T T_R d \leq \epsilon'$. Finally, (C5) is a linear equality constraint.
The problem~\eqref{eqn:alternative_problem} then reduces to
\begin{align*}
\underset{d}{\text{maximize } }\hspace{4mm} & T_Z(i,:) d   \\
\text{subject to } \hspace{4mm} 
  &||Q d||_{\infty} \leq 1 \hspace{5mm}d^T T_R^T T_R d \leq \epsilon'\hspace{5mm}Fd=0. 
     \end{align*} 
This is a problem of maximizing a concave function under a convex symmetric inequality constraints, and linear equality constraints, so it is convex.\hfill $\square$

Note that if there exists $d$ such that $T_zd\neq 0$ and $[Q^T\hspace{1mm}T^T_R \hspace{1mm}F^T]^Td=0$, and if $\epsilon' \geq 0$, the problem~\eqref{eqn:alternative_problem} is unbounded for at least one $i$. 
The attacker can then make the deterministic part of a critical state arbitrary large, so the influence of the stochastic components becomes negligible. 
In that case, the optimal value of Problem~1 (resp. Problem~2) goes to 1 (resp. $+\infty$). 
This special case can be checked separately, so we exclude it from the further analysis.

\begin{assumption} \label{assumption:null_spaces}
$\text{null} \hspace{1mm}([Q^T\hspace{1mm}T^T_R \hspace{1mm}F^T]^T) \subseteq \text{null} (\hspace{1mm}T_Z).  $
\end{assumption}

Finally, we introduce the lemma that establishes a link between maximizing $\mathbb{E}\{ z_{1:N}^{(i)} \}$ and $\mathbb{P}(|z_{1:N}^{(i)}| > 1)$, which is used in the proof of Theorem~1. 
We also revisit Jensen's inequality, which is used in the proof of Theorem~2.

\begin{lemma}
\label{lem:impact_metric_properties}
Let $\mu_i(d)=\mathbb{E}\{ z_{1:N}^{(i)} \}$, $P_i(d)=\mathbb{P}(|z_{1:N}^{(i)}| > 1)$, and $\mathcal{C}_d$ be a symmetric set. If $d^*_1 = \text{argmax}_{d\in \mathcal{C}_d} \hspace{1mm} \mu_i(d)$ and $d^*_2 = \text{argmax}_{d\in \mathcal{C}_d} \hspace{1mm} P_i(d)$, then $P_i(d^*_1)=P_i(d^*_2)$. 
\end{lemma}

\begin{lemma}
\label{lem:jensen_im}
(Jensen's inequality~\cite{PERLMAN197452}) Let $\phi$ be a convex function defined on a convex subset $\mathcal{C}_\phi$ of $\mathbb{R}^n$, and let $X$ be an $n$-dimensional integrable random vector that satisfies $\mathbb{P}(X\in \mathcal{C}_\phi)=1$. Then $\phi(\mathbb{E}\{X\})\leq \mathbb{E}\{\phi(X)\}. $
\end{lemma}

\subsection{Solving Problem~\ref{problem:impact_estimation} }

We now introduce Algorithm~\ref{algorithm:impact_calculation}, and prove that it solves Problem~\ref{problem:impact_estimation}. 
This algorithm solves~\eqref{eqn:alternative_problem} for each $i$ from $\{1,\ldots,n_zN\}$. 
Once the solution $d_i^{*}$ is obtained, the algorithm calculates 
$P_i(d_i^{*})=\mathbb{P}(|z^{(i)}_{1:N}| > 1)$. 
This is possible since $z_{1:N}\sim\mathcal{N}(T_Z d_i^{*},\Sigma_{Z})$ (Lemma~\ref{lem:critical_states_properties}), so 
$z_{1:N}^{(i)}(N)$ is Gaussian random variable with distribution  $\mathcal{N}(T_Z(i,:) d_i^{*},\Sigma_{Z}(i,i))$. 
The largest $P_i(d_i^{*})$ is then returned as the attack impact.


\begin{theorem} \label{theorem:solving_problem_1}
Let $I_{1}^*$ be the optimal value of Problem~\ref{problem:impact_estimation} and let $I'_1$ be the value returned by  Algorithm~\ref{algorithm:impact_calculation}. Then $I_{1}^*=I'_1$.
\end{theorem}
\textit{Proof. } 
The constraints of Problem~\ref{problem:impact_estimation} are independent of $i$. 
Thus, Problem~\ref{problem:impact_estimation} can be solved in the following two steps.
In the first step, we solve  $P_i^{*}=\text{maximize}_{d\in \mathcal{C}} \hspace{1mm} \mathbb{P}(|z^{(i)}_{0:N}| > 1)$ 
for each $i\in\{1,\ldots,n_zN\}$. In the second step, we find the optimal value of the attack impact as 
$I_{1}^*=\max_{i \in \{1,\ldots,n_zN\}}P_i^{*}.$

We now show that Algorithm~\ref{algorithm:impact_calculation} performs these two steps. 
Firstly, Algorithm~\ref{algorithm:impact_calculation} finds $d^*_i$ that maximizes $\mu_i(d)$ over $\mathcal{C}$ for each $i$. 
Recall that $\mathcal{C}$ is convex and symmetric constraint in $d$ (Proposition~\ref{theorem:convexity_problem_2}).
Therefore, based on Lemma~\ref{lem:impact_metric_properties}, $d^*_i$ is also maximizer of $\mathbb{P}(|z^{(i)}_{0:N}| > 1)$ over $\mathcal{C}$.
Thus, the algorithm finds $P_i^{*}$ for each $i$, which is the first step of the procedure. 
The algorithm then performs the second step of the procedure (Line~7). 
Thus, Algorithm~\ref{algorithm:impact_calculation} returns the optimal value of Problem~\ref{problem:impact_estimation}, which concludes the proof.    \hfill $\square$
\begin{algorithm} [t!]
\caption{Calculating attack impact}
\begin{algorithmic}[1]
\State \textbf{Input:} Model~\eqref{eqn:system_extended_eq_at}, $Q_{y_r}$, $Q_{z}$, $F_a$, $T_{sx}$, $T_{sr}$, $T_{sf}$, $N$, $\epsilon$
\State \textbf{Output:} $I'_1$
\For {$i=1:1:n_zN$}
\State Solve~\eqref{eqn:alternative_problem} for ${z_{1:N}^{(i)}}$ to obtain  $d_i^*$
\State Calculate $P_i(d_i^{*})=\mathbb{P}(|z_{1:N}^{(i)}| > 1)$
\EndFor
\State $I'_1= \max_{i\in\{1,\ldots,Nn_z\}}P_i(d_i^{*})$
\end{algorithmic}
\label{algorithm:impact_calculation}
\end{algorithm}

Theorem~\ref{theorem:solving_problem_1} represents an interesting extension of the work on deterministic systems~\cite{david}, which used the infinity norm based metric~\eqref{eqn:deterministic_metric}. 
 In particular, Theorem~\ref{theorem:solving_problem_1} shows that for the class of attacks we consider, the optimal value for the metric $I_1$ can be obtained by solving a set of convex optimization problems.
 This is the same property that $I$ had in the deterministic systems case.

\begin{remark}
In Algorithm~\ref{algorithm:impact_calculation}, the problem~\eqref{eqn:alternative_problem} needs to be solved $n_z N$ times. However, a crucial observation is that these problems are independent, so we can solve them in parallel.
Moreover, since we are performing off-line analysis, the execution time is not of critical importance. \hfill $\square$
\end{remark}
\begin{remark}
Calculating $P_i(d_i^{*})$ needs to be done numerically. However, due to importance of scalar Gaussian distribution, this probability can be calculated with sufficiently large accuracy and in the computational time that is negligible compared to the computational time of solving~\eqref{eqn:alternative_problem}.  \hfill $\square$
\end{remark}
\subsection{Lower Bound for Problem~\ref{problem:impact_estimation_2}}

We now show that in the process of solving Problem~\ref{problem:impact_estimation}, we also obtain a lower bound for Problem~\ref{problem:impact_estimation_2}. 
Note that Algorithm~\ref{algorithm:impact_calculation} finds the optimal value $\mu_i(d_i^*)$ of~\eqref{eqn:alternative_problem} for each $i$ from $\{1,\ldots,Nn_z\}$. 
Let $I'_2$ be defined as  
\begin{equation}\label{eqn:lowerbound}
I'_2= \text{max}_{i \in \{1,\ldots,Nn_z\}} \hspace{1mm}\mu_i(d_i^*). 
\end{equation}
In what follows, we prove that $I'_2$ can be used as the aforementioned lower bound. 

\begin{theorem} \label{prop:alternative_impact_metric}
Let $I_{2}^*$ be the optimal value of Problem~\ref{problem:impact_estimation_2}. Then $I'_2 \leq I_{2}^*$, where $I'_2$ is given by~\eqref{eqn:lowerbound}.   
  \end{theorem}
  \textit{Proof. } Let $I''_2=\text{maximize}_{d \in \mathcal{C}} \hspace{1mm}  ||\mathbb{E}\{z_{1:N}\}||_{\infty}.$ 
We first show $I_2''=I_2'$.
 Since, $\mathbb{E}\{z_{1:N}\}=T_Z d$, then
\begin{align} 
I'_2&=\text{max}_{i \in \{1,\ldots,Nn_z\}} \text{maximize}_{d \in \mathcal{C}} \hspace{1mm} T_Z(i,:) d\label{eqref:prob_inf_norm_1}\\
I''_2&=\text{maximize}_{d \in \mathcal{C}} \hspace{1mm}  ||T_Z d||_{\infty}.\label{eqref:prob_inf_norm_2}
\end{align}
In our previous work~\cite{jezdimir_ECC}, we used the fact that the problem of maximizing  $||T_Z d||_{\infty}$ under the convex symmetric inequality constraints can be solved by maximizing $T_Z(i,:)d$ over that same constraints, and then finding maximum over $i$.
The same proof can be used for general symmetric convex constraints. 
We include the proof here for reader's convenience and for the sake of completeness. 

Let $d''_2$ be an optimal solution for which $I''_2$ is obtained, and let rewrite $I_2''$ as $I_2''=||T_Z d''_2||_\infty=|T_Z{(i'',:)}d''_2|$,  
where $i''$ is the number of row of the matrix $T_Z$ for which the maximum of $||T_Z d''_2||_\infty$ is achieved.
Thus $|T_Z(i'',:)d''_2| \geq T_Z(i,:)d$ for every $i$ from $\{1,\ldots,Nn_z\}$, and every $d$ from $\mathcal{C}$. 
Since the constraint on $d$ is $\mathcal{C}$ in both~\eqref{eqref:prob_inf_norm_1} and~\eqref{eqref:prob_inf_norm_2}, then $I'_2 \leq I_2''$. 
We now show that $I'_2 < I_2''$ is not possible by using the contradiction argument. 
Since $\mathcal{C}$ is symmetric constraint, we have that both $d''_2$ and $-d''_2$ are feasible points for~\eqref{eqref:prob_inf_norm_1} and~\eqref{eqref:prob_inf_norm_2}.
Moreover, either $T_Z(i'',:)d''_2$ or $T_Z(i'',:)(-d''_2)$ must be greater than $0$. 
Then it follows $\max\{-T_Z(i'',:)d_2'', T_Z(i'',:)d''_2\}=I''_2>I'_2.$ 
This contradicts the assumption that $I'_2$ is the optimal value of~\eqref{eqref:prob_inf_norm_1}, and we conclude $I'_2=I_2''$. 

We now use Jensen's inequality to finalize the proof. Let $X_z\sim \mathcal{N}(T_Z d''_2,\Sigma_Z)$, and note that this random vector is with finite mean (integrable) once Assumption~\ref{assumption:null_spaces} holds.
Since every norm is convex, it follows from Jensen's inequality
$$I'_2=I''_2=||\mathbb{E}\{X_z\}||_\infty \leq \mathbb{E}\{||X_z||_\infty\}=I_2(d''_2).$$
 Note that $d''_2$ is a feasible point for Problem~\ref{problem:impact_estimation_2}. 
Thus, we have $I'_2 \leq I_2(d''_2) \leq I_{2}^*$, since $I_{2}^*$ is the optimal value of Problem~\ref{problem:impact_estimation_2}. The proof is now completed. \hfill $\square$


\section{Attack Strategies Compatible with Problem~\ref{problem:impact_estimation}}\label{section:attack_strategies}

We now introduce attack strategies for which the impact estimation problem can be reduced to Problem~\ref{problem:impact_estimation}. 
To show compatibility, we need to verify if a strategy can be modeled by~\eqref{eqn:attack}, and if $a$ and $a_s$ satisfy~\eqref{eqn:constraint_a}--\eqref{eqn:as}.

\subsection{DoS, Rerouting, and Sign Alternation Attacks}

We first consider strategies that can be modeled by
\begin{equation}\label{eqn:DOS_strategy}
\tilde{y}(k)=\Lambda_{y} y(k)\hspace{15mm} \tilde{u}(k)=\Lambda_{u} u(k).
\end{equation}
Note that~\eqref{eqn:DOS_strategy} can be reduced to~\eqref{eqn:attack} by enforcing $a_y$, $a_u$ and $a_s$ to be zero. 
This constraint on $a_y$ and $a_u$ can be reduced to the constraint~\eqref{eqn:constraint_a}, 
by adopting $F_a$ to be the identity matrix. 
Further, if we set $T_{sx}$, $T_{sf}$, and $T_{sr}$ to zero, the stochastic part $a_s$ satisfies the constraint~\eqref{eqn:as}.
Therefore, the attack strategies that satisfy~\eqref{eqn:DOS_strategy} are according to equations~\eqref{eqn:attack}, \eqref{eqn:constraint_a}--\eqref{eqn:as}, so their impact can be obtained by solving Problem~\ref{problem:impact_estimation}. 
We now introduce three strategies that can be modeled by~\eqref{eqn:DOS_strategy}. 

The first strategy are DoS attacks. 
In this strategy, the attacker blocks the signals of the sensors $\bar{\mathcal{Y}}=\{j_1,\ldots,j_{n_{a_y}}\}$ and actuators         $\bar{\mathcal{U}}=\{i_1,\ldots,i_{n_{a_u}}\}$ from reaching their destination. 
For example, the attacker can physically damage these devices, or jam the network~\cite{cardenas2008research}.
The DoS attacks can be modeled by~\eqref{eqn:DOS_strategy} (see~\cite{cardenas2008research,amin2009safe}), where the matrices $\Lambda_{y}$ and $\Lambda_{u}$ are diagonal matrices defined by
\begin{equation}\label{eqn:DOS_matrices}
\Lambda_{y}(i,i)=\begin{cases}
1, \text{$i \notin \bar{\mathcal{Y}}$}\\
0,  \text{$i \in \bar{\mathcal{Y}}$}
\end{cases} 
\hspace{5mm}
\Lambda_{u}(i,i)=
\begin{cases}
1, \text{$i \notin \bar{\mathcal{U}}$ }\\
0, \text{$i \in \bar{\mathcal{U}}$}.
\end{cases}
\end{equation}

The second strategy are rerouting attacks. 
This strategy consists of the attacker permuting the values of the measurement and control signals under its control~\cite{ferrari2017detection,teixeira2015voltage}. 
The attacker can perform this attack by physically re-wiring the sensor cables, or by modifying the sender's address~\cite{ferrari2017detection}.
The measurements and control inputs during the rerouting attack are given by~\eqref{eqn:DOS_strategy}, where $\Lambda_{y}$ and $\Lambda_{u}$ are permutation matrices that satisfy $\Lambda_{y}(i,i)=1$ for $i \notin \bar{\mathcal{Y}}$ and $\Lambda_{u}(i,i)=1$ for $i \notin \bar{\mathcal{U}}$.

The third attack strategy are the sign alternation attacks, where the attacker flips the sign of the measurement and control signals under its control. 
Although simple, this attack can turn negative feedback into positive, and potentially destabilize the system. 
Moreover, it was shown that in certain configurations, this attack strategy leads to strictly stealthy attacks~\cite{7470566}. 
The attack is according to~\eqref{eqn:DOS_strategy}, where $\Lambda_{u}$ and $\Lambda_{y}$ are diagonal matrices given by
\begin{equation*}
\Lambda_{y}(i,i)=
\begin{cases}
-1, \hspace{2mm} \text{$i \in \bar{\mathcal{Y}}$}  \\
\hspace{2.5mm}1,\hspace{2mm} \text{$i \notin \bar{\mathcal{Y}}$} 
\end{cases}
 \hspace{5mm} 
\Lambda_{u}(i,i)=
\begin{cases}
-1, \hspace{2mm}\text{$i \in \bar{\mathcal{U}}$} \\
\hspace{2.5mm}1, \hspace{2mm} \text{$i \notin \bar{\mathcal{U}}$}.
\end{cases}
\end{equation*}


\subsection{FDI and Bias Injection Attacks}

In FDI attacks, the attacker is able to manipulate sensors $\bar{\mathcal{Y}}$ and actuators $\bar{\mathcal{U}}$, and knows the whole model of the plant~\cite{5718158,david}. Based on these resources, the attacker constructs an optimal attack sequence $a_{0:N}$ that maximizes some impact metric. Signals $\tilde{y}$ and $\tilde{u}$ are given by
\begin{equation}\label{eqn:FDI_jednacija}
 \tilde{y}(k)= y(k)+\Gamma_{y} a_y(k) \hspace{10mm} \tilde{u}(k)= u(k)+\Gamma_{u} a_u(k). 
 \end{equation}
The matrices $\Gamma_{y}$ and $\Gamma_{u}$ are defined based on $\mathcal{\bar{Y}}$ and $\mathcal{\bar{U}}$. 
If $\mathcal{\bar{Y}}=\{j_1,\ldots,j_{n_{a_y}}\}$, then the elements $(j_1,1), \ldots, (j_{n_{a_y}},n_{a_y})$ of the matrix $\Gamma_{y}$ are equal to one, and the remaining elements are equal to zero. 
The matrix $\Gamma_{u}$ is defined in a similar manner based on $\bar{\mathcal{U}}$. 
Note that~\eqref{eqn:FDI_jednacija} can be reduced to~\eqref{eqn:attack} by adopting $\Lambda_{y}=\textbf{I}_{n_y}$ and $\Lambda_{u}=\textbf{I}_{n_u}$, and enforcing $a_s=0$. 
We do not have any restrictions on $a_u$ and $a_y$, so we select $F_a$ to be the zero matrix in the constraint~\eqref{eqn:constraint_a}. 
The stochastic part of the attack is $a_s=0$, which satisfies the constraint of the form~\eqref{eqn:as}.
Therefore, this strategy satisfies the attack model we proposed, and its impact can be estimated by solving Problem~\ref{problem:impact_estimation}. 

\begin{remark}
If $N\rightarrow +\infty$ and $\epsilon = 0$, it can be shown that Problem~\ref{problem:impact_estimation} reduces to
\begin{align*}
\underset{y_r,a_0,\ldots}{\text{maximize}} \hspace{1mm}&I_1(y_r,a_0,\ldots)\\
\text{subject to} \hspace{1mm} &x_e(k+1)=\tilde{A} x_e(k)+ \tilde{G} a(k)  &&x_e(0)=T_{0} y_r \\
    &\hspace{12.5mm}0=\tilde{C} x_e(k)+ \tilde{H} a(k) &&||Q_{y_r} y_r||_{\infty} \leq 1.
\end{align*}
What is interesting to note is that the constraints enforce zero dynamics attacks, with the initial condition restricted to $T_{0} y_r$~\cite{bias,6545301}.
If we additionally impose $y_r=0$, then perfectly undetectable attacks are enforced~\cite{6816560,7479526}.  \hfill $\square$
\end{remark}

We now introduce the bias injection attacks. 
This strategy is a less sophisticated type of injection attacks, since the attacker adds a constant bias to the measurement and control signals under its control~\cite{ahmed2017model,bias}. 
Signals $\tilde{y}$ and $\tilde{u}$ in this attack strategy are modeled same as in~\eqref{eqn:FDI_jednacija}. 
In addition to the constraints for FDI attacks, the deterministic parts of the attack need to satisfy $a_y(k)=a_y(0)$, $a_u(k)=a_u(0)$ for $N \geq k > 0$. 
These constraints on $a_u$ and $a_y$ are reducible to~\eqref{eqn:constraint_a}. 
 Thus, this strategy is  also compatible with our framework. 

One can also imagine the combination of FDI and DoS attacks~\cite{7799046,7778773},  where the attacker can inject corrupted data to the measurements $\bar{\mathcal{Y}}_{I}$ and controls $\bar{\mathcal{U}}_{I}$, but can only deny access to $\bar{\mathcal{Y}}_{D}$ and $\bar{\mathcal{U}}_{D}$. The attack can then be modeled as
\begin{equation*}
\tilde{y}(k)= \Lambda_{y}  y(k)+\Gamma_{y} a_y(k)  \hspace{7mm} \tilde{u}(k)=\Lambda_{u} u(k)+\Gamma_{u} a_u(k)
\end{equation*}
where $\Lambda_{y}$ and $\Lambda_{u}$ are defined based on $\bar{\mathcal{Y}}_{D}$ and $\bar{\mathcal{U}}_{D}$ as in~\eqref{eqn:DOS_matrices}, and $\Gamma_{y}$ and $\Gamma_{u}$ are defined based on $\bar{\mathcal{Y}}_{I}$ and $\bar{\mathcal{U}}_{I}$ as in~\eqref{eqn:FDI_jednacija}. 
This equation is of the same form as~\eqref{eqn:attack}, with $a_s=0$. The signals $a_y$ and $a_u$ are free for the attacker to choose, which is according to~\eqref{eqn:constraint_a}. Since $a_s=0$,~\eqref{eqn:as} is also satisfied. Thus, we conclude that this strategy is compatible with our framework. 

\subsection{Replay Attacks}
Replay attacks are inspired by the well known Stuxnet malware~\cite{stuxnet}. The attacker first records steady state sensor measurements. It then starts sending these healthy measurements from the sensors $\bar{\mathcal{Y}}$ under its control, while at the same time manipulates the compromised actuators $\bar{\mathcal{U}}$ in a malicious way. The replay attack on sensors can be modeled by
\begin{align}\label{eqn:measurement_replay}
\tilde{y}(k)&= \Lambda_{y}  y(k)+\Gamma_{y} a_s(k) 
\end{align}
where $\Lambda_{y}$ is given as in~\eqref{eqn:DOS_matrices}. The signal $a_s$ is given by 
\begin{align}\label{eqn:stoch_attack_replay}
a_s(k)= C_{\bar{\mathcal{Y}}}y(k-N-1)
\end{align}
where $C_{\bar{\mathcal{Y}}} \in \mathbb{R}^{n_{a_y} \times n_y}$ is defined based on  the set of attacked sensors $\bar{\mathcal{Y}}$. If $\bar{\mathcal{Y}}=\{j_1,\ldots,j_{n_{a_y}}\}$, then the elements $(1,j_1), \ldots, (n_{a_y},j_{n_{a_y}})$ of $C_{\bar{\mathcal{Y}}}$ are one, and the remaining elements are zero. 
In other words, the attacker replaces the attacked measurements with the measurements of the normal operation previously recorded at times $k\in[-N-1,-1]$. 

The signal $a_u$ sent to the actuators could be modeled in different ways. For example, in our previous work~\cite{jezdimir_ECC}, we assumed that $a_u$ is a constant bias, that is
\begin{align} \label{eqn:control_replay_1}
\tilde{u}(k)= u(k)+\Gamma_{u}a_u(0)
 \end{align} 
where $\Gamma_{u}$ is defined as in~\eqref{eqn:FDI_jednacija}. This models the case where the attacker sends some large signal to actuators.  
Another scenario can be a DoS attack against the actuators
\begin{align}\label{eqn:control_replay_2}
\tilde{u}(k)= \Lambda_{u} u(k)
 \end{align} 
where $\Lambda_{u}$ is given as in~\eqref{eqn:DOS_strategy}. 
Both of the previously introduced scenarios of replay attacks can be captured with the general attack model. 
The proof is available in Appendix~\ref{appendix:replay}.

\section{Numerical Example}\label{section:simulations}

We now illustrate how the modeling framework we proposed can be used for risk assessment, and explore how the tuning parameters influence the impact of different strategies.

\subsection{System Model}
\begin{figure}
    \centering
  \includegraphics{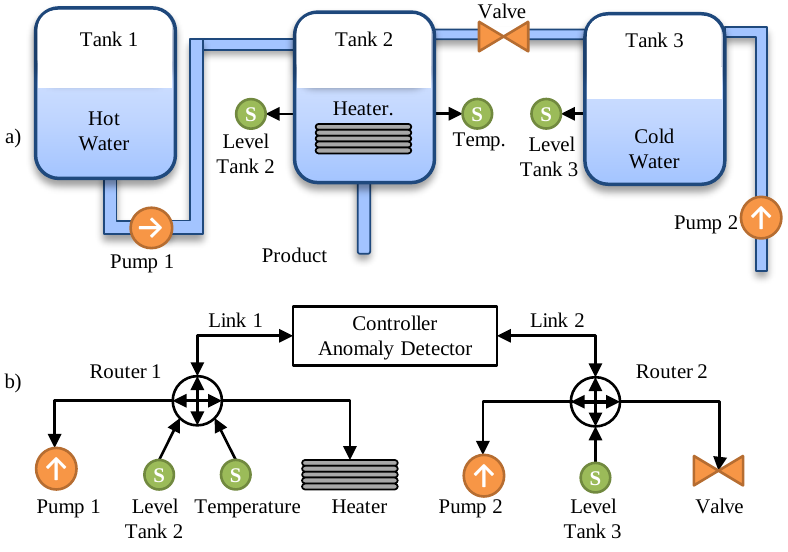}
  \caption{a) Physical part of chemical process with four actuators (two pumps, heater, and valve), and three sensors (two level sensors and one temperature sensor); b) Cyber infrastructure of the process.   }
  \label{figure:Tank_Physical}
\end{figure}

We consider a chemical process from~\cite{blanke2006diagnosis} shown in Figure~\ref{figure:Tank_Physical}~a).  
The states are  the volume in Tank~3 ($x_1$), the volume in Tank~2 ($x_2$), and the temperature in Tank~2 ($x_3$). 
The control signals are the flow rate of Pump~2 ($u_1$), the openness of the valve ($u_2$), the flow rate of Pump~1 ($u_3$), and the power of the heater ($u_4$). 
We assume that the control objective is to keep a constant temperature in Tank 2. 
The objective is achieved by injecting hot water from Tank 1, and cold water from Tank 3. 

The plant is described by 
\begin{equation*}
\small
\begin{aligned}
A&=\begin{bmatrix}
0.96 &  0 &  0              \\
0.04 & 0.97&  0               \\
-0.04 & 0 & 0.90               \\
\end{bmatrix}
B=\begin{bmatrix}
8.8 & -2.3 & 0 & 0               \\
0.20 & 2.2& 4.9   &    0        \\
-0.21 & -2.2 & 1.9 &    21           \\
\end{bmatrix}
\end{aligned}
\end{equation*}
$C=\textbf{I}_3$, $\Sigma_v=0.05\hspace{0.5mm}\textbf{I}_3$, and $\Sigma_w=0.01\hspace{0.5mm}\textbf{I}_3$. 
The matrices of the controller are given by 
\begin{equation*}
\small
\begin{aligned}
L_{\hat{x}}&=0.01\begin{bmatrix} 
 10   & 1.8   &-0.1 \\
   -2.0  &  7.1  & -0.5\\
    1.4  & 16  &  0.2 \\
   -0.4 &  -0.7  &  4.2\\
\end{bmatrix} 
L_{y_r}=0.01\begin{bmatrix}
 11&   11   & 0 \\
   -1  & 44   & 0 \\
         0      &   0        & 0           \\
    0    & 4.7    &4.7 \\
\end{bmatrix}.
\end{aligned}
\end{equation*}
We assumed $Q_{y_r}=0.4\hspace{0.5mm}\textbf{I}_3$, and we used the steady state Kalman filter as an estimator.

  A cyber-infrastructure is assumed to be as shown in Figure~\ref{figure:Tank_Physical}~b).  
It was identified that the communication link between Router~1 and the controller is not protected (Vulnerability~1).
The same holds for the communication link between Router~2 and the controller (Vulnerability~2). 
If the attacker exploits Vulnerability~1, it gains control over sensors $y_2,y_3$ and actuators $u_3,u_4$.
In the case of Vulnerability~2, sensor $y_1$ and actuators $u_1,u_2$ can be manipulated.

\subsection{Risk Assessment}
Our goal is to investigate which of these two vulnerabilities is more threatening to our system. 
For this, we use our framework. 
We set $Q_z=[\textbf{0}_{1 \times 2} \hspace{2mm}  1/3 \hspace{2mm} \textbf{0}_{1 \times 3}]$, $N=10$, $\epsilon=0.3$, and we use the metric $I_1$. 
We then calculated the impacts of DoS~\cite{cardenas2008research,amin2009safe}, rerouting~\cite{teixeira2015voltage,ferrari2017detection}, replay~\cite{6612700}, FDI~\cite{5718158,david} and bias injection~\cite{ahmed2017model,bias} attacks once Vulnerability~1 and Vulnerability~2 are exploited. 
Since the attacker can conduct DoS and rerouting attacks in multiple ways, we calculated the worst case impact for these strategies. 
For the replay strategy, the attack against the actuators was modeled according to~\eqref{eqn:control_replay_2}.

    \begin{figure}
   \centering
  \includegraphics[width=75mm]{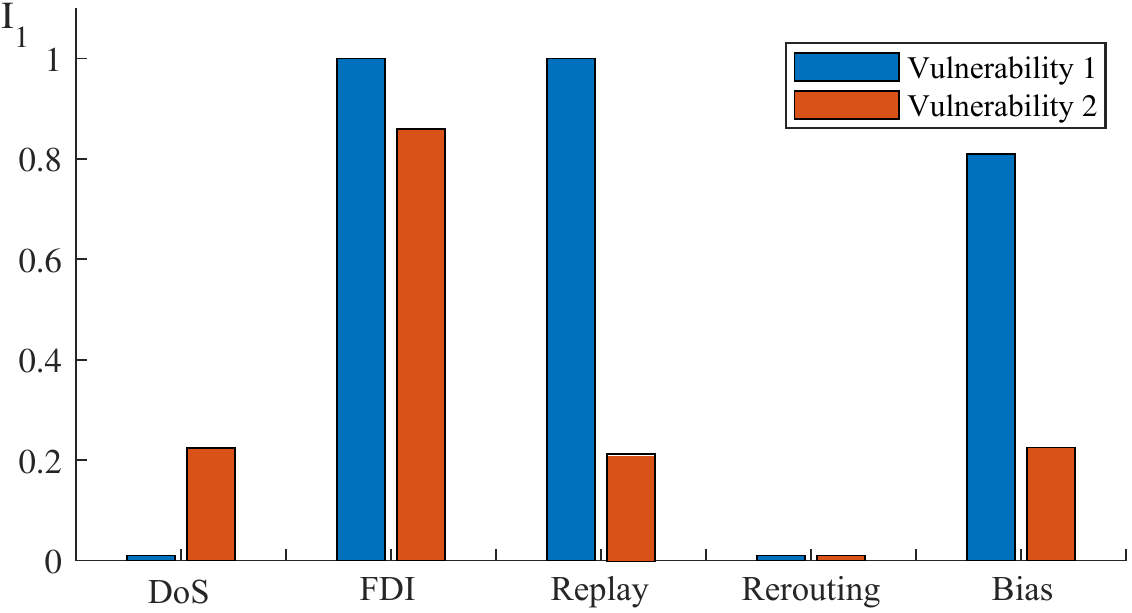}
  \caption{Impact of different attacks once Vulnerability~1 and Vulnerability~2 are exploited. }
  \label{figure:impact_analysis}
\end{figure}
    \begin{figure}   
    \centering
  \includegraphics[width=75mm]{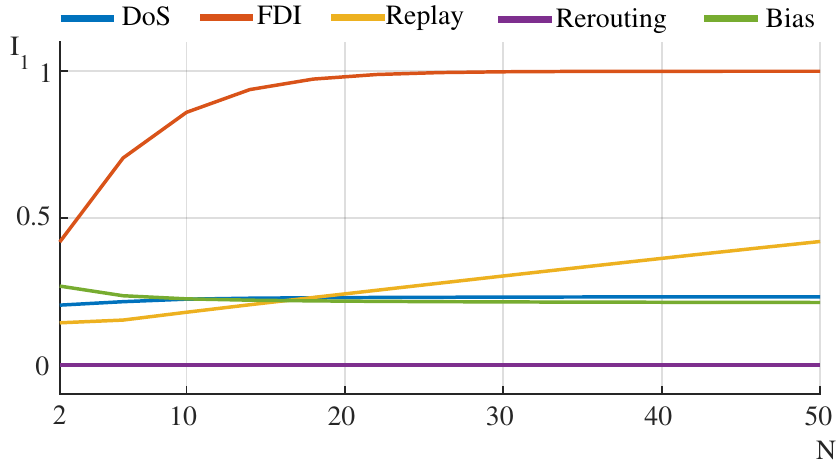}
  \caption{Impact of different attack strategies in respect to $N$. }
  \label{figure:impact_N}
\end{figure}

The results of the analysis are illustrated in Figure~\ref{figure:impact_analysis}. 
Firstly, note that the impact of different strategies may result in different conclusions concerning the importance of vulnerabilities. 
For instance, based on the impact of DoS attacks, it follows that Vulnerability~2 is more important to be prevented than Vulnerability~1. 
On the other hand, based on the impact of replay, FDI, and bias attacks, Vulnerability~1 is more critical.
The impact of rerouting attacks was not informative, since it was equal to 0 in both of the cases.
This illustrates the point that in some cases, attack impact of several strategies needs to be taken into consideration to decide on importance of vulnerabilities. 
In this case, we can give higher priority to Vulnerability~1, since the impact of majority of attack strategies is larger for this vulnerability. 

It is also interesting to note that sometimes less complex attacks can be just as dangerous as full model knowledge FDI attacks.
For example, in the case of Vulnerability~1, replay attacks lead to the same impact as FDI attacks.
Finally, we note that the stochastic model of the system can considerably influence the impact of some attacks. 
In particular, rerouting attacks proved to be harmless in this framework, because they were detectable in both of the scenarios. 
However, in our previous study on deterministic systems~\cite{jezdimir_ECC}, these attacks had impact comparable with DoS and bias injection attacks.  

\subsection{Tuning Parameters}

In our framework, the impact estimation problem has $\epsilon$ and $N$ as tunning parameters. 
By increasing $\epsilon$, the stealthiness constraint becomes easier to satisfy.
Thus, it is clear that the impact is nondecreasing with respect to $\epsilon$. 
However, the connection of the impact to the length of horizon $N$ is not obvious.
To illustrate some interesting facts, we investigate how the impact changes when we vary $N$ in the range $2$ to $50$. 
We fixed other modeling parameters to be the same as in the previous two sections, assumed Vulnerability 2 to be exploited, and considered the same attack strategies. 

A plot of the impact of different strategies in respect to $N$ is shown on Figure~\ref{figure:impact_N}.
The first interesting observation is that the impact of almost all the strategies seems to converge to a steady state relatively quickly. 
In fact, only the impact of the replay strategy keeps increasing over time. 
The second observation is that attack impact can also be decreasing with $N$, as in the case of bias injection attacks.
We find the reason to be that the stealthiness constraint becomes harder to satisfy, while the number of decision variables in the problem remains the same. 
Both of these observations point out that in certain cases, we do not have to consider long horizon lengths in order to find the worst case attack impact.  


\section{Conclusion and Future Work}\label  {section:conclusion}

We proposed a framework for estimating impact of a range of cyber-attack strategies, which is independent of the choice of anomaly detector. 
Furthermore, we suggested two alternatives for the impact metric based on the infinity norm that can be used in stochastic systems. 
For the first metric, we proved that the problem of estimating attack impact can be reduced to the problem of solving a set of convex problems, so the exact solution can be obtained efficiently (Theorem~1). 
For the second metric, an easily computable lower bound was proposed (Theorem~2).  
Finally, we demonstrated how our modeling framework can be used for risk assessment on an illustrative example.
There are two main directions that we plan to explore in
the future work. Firstly, we plan to further study monotonicity of the attack impact with respect to $N$. Secondly, we will investigate if the impact of feedback attacks can be analyzed using the proposed framework. 


\appendix

\subsection{Proofs of the Lemmas from Section~\ref{section:main_results}.A }

\textit{Proof of Lemma~\ref{lem:lemma_initial}.} In absence of attacks, the extended state $x_e$ is given by
\begin{align}\label{eqn:non_attacked_matrices}
x_e(k+1)=A_e x_e(k)+B_e f(k)+E_e y_r
\end{align}
where $A_e$,$B_e$ and $E_e$ are respectively obtained from $\tilde{A}$, $\tilde{B}$, and $\tilde{E}$ by setting $\Lambda_y=\textbf{I}_{n_y}$ and $\Lambda_u=\textbf{I}_{n_u}$, and $f$ is zero mean Gaussian white process whose covariance we denote by $\Sigma_f$. 

Assume $y_r=0$. 
Since $A_e$ is assumed to be asymptotically stable, it is known that $x_e$ is zero mean Gaussian stationary process with the covariance matrix obtained as the solution of the Lyapunov equation $\Sigma_{0}=A_e\Sigma_{0}A_e^T+B_e \Sigma_f B_{e}^T$ (see~\cite[Chapter 4]{optimalfiltering}). 
Once $y_r\neq0$, only the mean value of the process changes. 
Since we assumed that the system has reached the steady state, we have $\mathbb{E}\{ x_e \}=A_e \mathbb{E}\{ x_e\}+E_ey_r$, which implies $\mathbb{E}\{ x_e \}=T_{0} y_r$, where $T_{0}=(\textbf{I}_{2n_x}-A_e)^{-1} E_e$. \hfill $\square$

\textit{Proof of Lemma~\ref{lem:critical_states_properties}.} We first consider $z_{1:N}$. Since the system before and after the attack is linear, $z_{1:N}$ can be expressed as
\begin{equation} \label{eqn:critical_states_properties_eq1}
z_{1:N}= P_x x_e(N_s)+P_f f_{N_s:N}+P_r y_r+P_a a_{0:N}+P_s a_{s{0:N}}. 
\end{equation}
From~\eqref{eqn:as}, it follows
\begin{equation}\label{eqn:critical_states_properties_eq2}
z_{1:N}= P'_x x_e(N_s)+P'_f f_{N_s:N}+P'_r y_r+P_a a_{0:N}
\end{equation}
where $P'_f=P_f+P_s[T_{sf} \hspace{2mm} \textbf{0}_{(N+1)n_{a_y}\times(N+1)n_f}]$, $P'_x= P_x+P_s T_{sx}$, and $P'_r=P_r+P_s T_{sr}$. Since $x_e(N_s)$ and $f_{N_s:N}$ are \textit{independent} Gaussian vectors, and $a_{0:N}$ and $y_r$ are deterministic, $z_{1:N}$ is a Gaussian vector. Using the linearity property of expected value operator and Lemma~\ref{lem:lemma_initial}, we get 
\begin{equation}
\mathbb{E}\{z_{1:N}\}=P'_x T_{0} y_r+P'_f 0 + P'_r y_r+  P_a a_{0:N}= T_Z d
\end{equation} 
where $T_Z=\begin{bmatrix} P_a & P'_x T_{0}+P'_r   \end{bmatrix}$. Let the covariance matrix of $f_{N_s:N}$ be denoted with $\Sigma_F$. We then have
\begin{equation}\label{eqn:critical_states_properties_eqlast}
\Sigma_{Z}= P'_x \Sigma_{0} (P'_x )^T+P'_f \Sigma_{F} (P'_f) ^T
\end{equation} 
where we used that $x_e(N_s)$ and $f_{N_s:N}$ are independent. 
Since $\Sigma_{0}$ and $\Sigma_{F}$ are dependent only on the noise properties and the system matrices, $\Sigma_{Z}$ is fixed and not dependent on $d$. 
The proof that $\tilde{r}_{0:N}\sim \mathcal{N}(T_Rd,\Sigma_{R})$ follows the same steps~\eqref{eqn:critical_states_properties_eq1}--\eqref{eqn:critical_states_properties_eqlast}, so we omit it due to page limit. 

It remains to prove that $\Sigma_{Z}$ is positive definite. Equation~\eqref{eqn:critical_states_properties_eq2} can be rewritten as 
\begin{multline*}
z_{1:N}= P'_x x_e(N_s)+P_{fp} f_{N_s:-1}+P_{v}v_{0:N-1}+P_{w}w_{0:N-1}\\+P'_r y_r+P_a a_{0:N}+0v(N)+0w(N).
\end{multline*}
Since $P'_x x_e(N_s)$, $P_{fp} f_{N_s:-1}$, $P_{v}v_{0:N-1}$, $P_{w}w_{0:N-1}$ are independent Gaussian vectors, $\Sigma_Z$ is the sum of covariance matrices of these vectors. Thus, it suffices to prove that one of these vectors has a positive definite covariance matrix. It can be shown from~\eqref{eqn:systemg} and $z(k)=Q_z x(k)$ that $P_v$ is the lower triangular matrix of the form 
$$\small P_v=\begin{bmatrix}Q_z & \textbf{0}_{n_z\times n_x} & \ldots &\textbf{0}_{n_z\times n_x} \\ \x & Q_z & \ldots &\textbf{0}_{n_z\times n_x} \\ \vdots & \vdots & \ddots & \vdots \\ \x & \x & \ldots &Q_z \end{bmatrix}. $$
It then follows that $\text{null}(P_v^T)=\emptyset$, since $Q_z$ has full row rank. Let $\Sigma_{V}$ be the covariance matrix of $v_{0:N-1}$. Since $\Sigma_{v}$ is positive definite, so it is $\Sigma_{V}$. 
Thus, $P_v \Sigma_{V} (P_v) ^T$ is positive definite, which implies that $\Sigma_Z$ is positive definite. \hfill $\square$

\textit{Proof of Lemma~\ref{lem:stealthiness_constraint_properties}. }Let $Y_1$ and $Y_2$ be random vectors with the distributions $\mathcal{N}(\mu_1,\Sigma_1)$ and $\mathcal{N}(\mu_2,\Sigma_2)$, respectively. Let $\Sigma_1$ and $\Sigma_2$ be positive definite. Then $$\mathcal{D}(Y_1||Y_2)=0.5(\text{tr} (\Sigma^{-1}_2\Sigma_1) + ||\mu_2-\mu_1||_{\Sigma_2^{-1}}+\text{ln}\frac{ \text{det}(\Sigma_2) }{\text{det}(\Sigma_1)}-n)$$
where $n$ is the dimension of $Y_1$ and $Y_2$~\cite{duchi2007derivations}.
In our case, the distributions of $\tilde{r}_{0:N}$ and $r_{0:N}$ are $\mathcal{N}(T_Rd,\Sigma_{R})$ and $\mathcal{N}(\textbf{0}_{(N+1)n_y \times 1},\textbf{I}_{(N+1)n_y})$, respectively. Thus, it follows
\begin{multline*}
\mathcal{D}(\tilde{r}_{0:N}||r_{0:N}) =0.5\big( \text{tr}(\Sigma_R) + (T_Rd)^T T_R d-(N+1)n_y -\\
\text{ln }\text{det}(\Sigma_R)  \big)=c_{KL}+\frac{1}{2}d^T T_R^TT_Rd
\end{multline*} 
where $c_{KL}=0.5(\text{tr}(\Sigma_R)-(N+1)n_y -\text{ln }\text{det}(\Sigma_R))$. 
Therefore,~\eqref{eqn:stealthiness} becomes $d^T T_R^T T_Rd \leq \epsilon'$, where $\epsilon'=2(\epsilon(N+1)-c_{KL})=(N+1)(2\epsilon+n_y)- \text{tr}(\Sigma_R) +\text{ln }\text{det}(\Sigma_R)$. \hfill $\square$


\textit{Proof of Lemma~\ref{lem:impact_metric_properties}.} From Lemma~\ref{lem:critical_states_properties},  $z_{1:N}$ is a Gaussian random vector with the non-degenerate distribution $\mathcal{N}(T_Z d,\Sigma_{Z})$. Thus, $z_{1:N}^{(i)}$ is a Gaussian random variable with the distribution $\mathcal{N}(\mu_i(d),\sigma_i^2)$, where $\mu_i(d)=T_Z(i,:) d$, and $\sigma_i^2=\Sigma_{Z}(i,i)$. 

We first consider probability $P'_i(d)=\mathbb{P}(|z_{1:N}^{(i)}| \leq 1)$.
  Let  $c_1=1/(\sqrt{2\pi}\sigma_i)$ and $c_2=-1/({2\sigma_i^2})$. We then have
\begin{equation} \label{eqn:integral_help}
 \small
\begin{aligned}
P'_i(d)=\int_{-1}^{1} c_1 e^{c_2(x-\mu_i(d))^2}dx =\int_{-1-\mu_i(d)}^{1-\mu_i(d)} c_1 e^{c_2 t^2}dt.
\end{aligned}
\end{equation}
Note that $d$ influences $P'_i(d)$ only through $\mu_i(d)$. In what follows, we outline two important properties of $P'_i(d)$ in respect to $\mu_i(d)$.

Firstly, note that $e^{-c_2 t^2}$ is symmetric in $t$, and the bounds of the integral~\eqref{eqn:integral_help} are $-1-\mu_i(d)$ and $1-\mu_i(d)$. Therefore, $P'_i(d)$ is symmetric in $\mu_i(d)$. Secondly, we show that $P_i'(d)$ is decreasing when $|\mu_i(d)|$ is increasing. 
Let $\mu_i(d_1)=\mu_i$ and $\mu_i(d_2)=\mu_i+\delta$, where $\mu_i \geq 0$ and $\delta > 0$. Then
\begin{equation*}
\small
\begin{aligned}
P'_i(d_1)&=\underbrace{\int^{1-\mu_i-\delta}_{-1-\mu_i} c_1e^{ c_2 t^2}dt}_{=I}
+\underbrace{\int^{1-\mu_i}_{1-\mu_i-\delta}c_1 e^{c_2t^2}dt}_{=\Delta_1} \\
P_i'(d_2)&=\underbrace{\int^{-1-\mu_i}_{-1-\mu_i-\delta}c_1 e^{c_2 t^2}dt}_{=\Delta_2}+\underbrace{\int^{1-\mu_i-\delta}_{-1-\mu_i} c_1 e^{c_2 t^2}dt}_{=I}.
\end{aligned}
\end{equation*}
Note that the intervals $$[1-\mu_i-\delta,1-\mu_i] \hspace{10mm}
[-1-\mu_i-\delta,-1-\mu_i]$$ are of the same length $\delta$. Additionally 
$$|1-\mu_i|\leq|-1-\mu_i|$$ 
and $e^{-c_2 t^2}$ is decreasing in $|t|$. Hence, we conclude $\Delta_2 < \Delta_1$. Therefore, $P_i'(d)$ is decreasing when $\mu_i(d) \geq 0$ is increasing. Additionally, because of symmetry of $P_i'(d)$, we have that $P_i'(d)$ is decreasing when $|\mu_i(d)|$ is increasing. 

From the previous discussion, we conclude that $P_i(d)$ is increasing when $|\mu_i(d)|$ is increasing, since $P_i(d)=1-P_i'(d)$.  
Furthermore, if we have a symmetric set $\mathcal{C}_d$, and if we find $d^*$ that maximizes $\mu_i(d)$, then we know that $d^*$ maximizes $|\mu_i(d)|$ over $\mathcal{C}_d$ as well. Finally, this $d^*$ maximizes $P_i(d)$ on $\mathcal{C}_d$, since $P_i(d)$ is increasing with $|\mu_i(d)|$. \hfill $\square$

%
%
\subsection{Replay Attacks}\label{appendix:replay}

Firstly, the equations~\eqref{eqn:measurement_replay}--\eqref{eqn:control_replay_2} are all according to~\eqref{eqn:attack}. 
The deterministic part $a_y$ satisfies~\eqref{eqn:constraint_a}, since $a_y=0$.
The same holds for $a_u$ if~\eqref{eqn:control_replay_2} is used to model actuator attacks. 
If~\eqref{eqn:control_replay_1} is used, $a_u$ needs to satisfy $a_u(k)=a_u(0)$ for $0 \leq k\leq N$, which is also reducible to~\eqref{eqn:constraint_a}.

It remains to be shown that the stochastic part of the attack $a_s$ satisfies~\eqref{eqn:as}. 
Let $N_s=-N-1$, and let $k$ be the time instant that satisfies $N_s \leq  k < 0$. From~\eqref{eqn:non_attacked_matrices}, $x_e(k)$ in the absence of attacks is dependent only on $x(N_s)$, $f$, and $y_r$, so $C_{\bar{\mathcal{Y}}}y(k)$ can be expressed as $$C_{\bar{\mathcal{Y}}}y(k)= T^{(k)}_{sx} x(N_s)+ T^{(k)}_{sr} y_r + T^{(k)}_{sf} f_{N_s:-1}. $$
Finally, from~\eqref{eqn:stoch_attack_replay}, $a_{s0:N}$ equals $$ a_{s0:N}= \begin{bmatrix}C_{\bar{\mathcal{Y}}}y(N_s) \\ \vdots \\ C_{\bar{\mathcal{Y}}}y(-1) \end{bmatrix}=T_{sx} x(N_s)+ T_{sr} y_r + T_{sf} f_{N_s:-1} $$ 
where  
$$T_{sx} =\begin{bmatrix}T^{(N_s)}_{sx} \\ \vdots \\ T^{(-1)}_{sx} \end{bmatrix} \hspace{5mm}  T_{sr} =\begin{bmatrix}T^{(N_s)}_{sr} \\ \vdots \\ T^{(-1)}_{sr} \end{bmatrix}\hspace{5mm}  T_{sf} =\begin{bmatrix}T^{(N_s)}_{sf} \\ \vdots \\ T^{(-1)}_{sf} \end{bmatrix}. $$

\balance

\bibliographystyle{IEEEtran}

\bibliography{autosam}

\end{document}